\input jytex.tex   
\typesize=10pt
\magnification=1200
\baselineskip17truept
\hsize=6truein\vsize=8.5truein
\sectionnumstyle{blank}
\chapternumstyle{blank}
\chapternum=1
\sectionnum=1
\pagenum=0

\def\begintitle{\pagenumstyle{blank}\parindent=0pt\begin{narrow}[0.4in]}
\def\endtitle{\end{narrow}\newpage\pagenumstyle{arabic}}


\def\beginexercise{\vskip 20truept\parindent=0pt\begin{narrow}[10 
truept]}
\def\endexercise{\vskip 10truept\end{narrow}}


\def\eql#1{\eqno\eqnlabel{#1}}
\def\ref{\reference}
\def\peq{\puteqn}
\def\pref{\putref}

\def\mgn{\marginnote}
\def\bex{\begin{exercise}}
\def\eex{\end{exercise}}


\def\mbox#1{{\leavevmode\hbox{#1}}}

\def\hspace#1{{\phantom{\mbox#1}}}

\def\al{\alpha}

\def\de{\delta}

\def\la{\lambda}
\def\La{\Lambda}
\def\om{\omega}

\def\vphi{\varphi}

\def\ze{\zeta}

\def\De{\Delta}

\def\zf{$\zeta$--function}
\def\zfs{$\zeta$--functions}


\def\frac#1/#2{\leavevmode\kern.1em
\raise.5ex\hbox{\the\scriptfont0 #1}\kern-.1em/\kern-.15em
\lower.25ex\hbox{\the\scriptfont0 #2}}
\def\sfrac#1/#2{\leavevmode\kern.1em
\raise.5ex\hbox{\the\scriptscriptfont0 #1}\kern-.1em/\kern-.15em
\lower.25ex\hbox{\the\scriptscriptfont0 #2}}

\def\gtorder{\mathrel{\raise.3ex\hbox{$>$}\mkern-14mu
             \lower0.6ex\hbox{$\sim$}}}
\def\ltorder{\mathrel{\raise.3ex\hbox{$<$}\mkern-14mu
             \lower0.6ex\hbox{$\sim$}}}

\def\semidirprod{\rlap{\ss C}\raise1pt\hbox{$\mkern.75mu\times$}}
\def\for{\lower6pt\hbox{$\Big|$}}
\def\fish{\kern-.25em{\phantom{abcde}\over \phantom{abcde}}\kern-.25em}


\def\boxit#1{\vbox{\hrule\hbox{\vrule\kern3pt
        \vbox{\kern3pt#1\kern3pt}\kern3pt\vrule}\hrule}}
\def\dalemb#1#2{{\vbox{\hrule height .#2pt
        \hbox{\vrule width.#2pt height#1pt \kern#1pt
                \vrule width.#2pt}
        \hrule height.#2pt}}}
\def\square{\mathord{\dalemb{5.9}{6}\hbox{\hskip1pt}}}

\def\frac#1#2{{{#1}\over{#2}}}


\def\cf{{\it cf }}


  %

\def\3j#1#2#3#4#5#6{\left\lgroup\matrix{#1&#2&#3\cr#4&#5&#6\cr}
\right\rgroup}

\def\man{{\cal M}}

\def\m?{\mgn{?}}

\def\beq{\begin{eqnarray}}
\def\eeq{\end{eqnarray}}


\def\aop#1#2#3{{\it Ann. Phys.} {\bf {#1}} (19{#2}) #3}

\def\cmp#1#2#3{{\it Comm. Math. Phys.} {\bf {#1}} (19{#2}) #3}
\def\cqg#1#2#3{{\it Class. Quant. Grav.} {\bf {#1}} (19{#2}) #3}

\def\jmp#1#2#3{{\it J. Math. Phys.} {\bf {#1}} (19{#2}) #3}
\def\jpa#1#2#3{{\it J. Phys.} {\bf A{#1}} (19{#2}) #3}
\def\mplA#1#2#3{{\it Mod. Phys. Lett.} {\bf A{#1}} (19{#2}) #3}
\def\np#1#2#3{{\it Nucl. Phys.} {\bf B{#1}} (19{#2}) #3}
\def\pl#1#2#3{{\it Phys. Lett.} {\bf {#1}} (19{#2}) #3}

\def\prD#1#2#3{{\it Phys. Rev.} {\bf D{#1}} (19{#2}) #3}

\def\cras#1#2#3{{\it Comptes Rend. Acad. Sci. (Paris)} {\bf{#1}} (#2) #3}

\def\mpcps#1#2#3{{\it Math. Proc. Camb. Phil. Soc.} {\bf{#1}} (19{#2}) #3}

\def\am#1#2#3{{\it Acta Mathematica} {\bf {#1}} (19{#2}) #3}
\def\aim#1#2#3{{\it Adv. in Math.} {\bf {#1}} (19{#2}) #3}
\def\ajm#1#2#3{{\it Am. J. Math.} {\bf {#1}} ({#2}) #3}

\def\aom#1#2#3{{\it Ann. of Math.} {\bf {#1}} (19{#2}) #3}

\def\cpde#1#2#3{{\it Comm. Partial Diff. Equns.} {\bf {#1}} (19{#2}) #3}

\def\invm#1#2#3{{\it Invent. Math.} {\bf {#1}} (19{#2}) #3}
\def\ijpam#1#2#3{{\it Ind. J. Pure and Appl. Math.} {\bf {#1}} (19{#2}) #3}
\def\jdg#1#2#3{{\it J. Diff. Geom.} {\bf {#1}} (19{#2}) #3}

\def\jmpa#1#2#3{{\it J. Math. Pures. Appl.} {\bf {#1}} ({#2}) #3}

\def\ojm#1#2#3{{\it Osaka J.Math.} {\bf {#1}} ({#2}) #3}

\def\pja#1#2#3{{\it Proc. Jap. Acad.} {\bf {A#1}} (19{#2}) #3}

\def\tams#1#2#3{{\it Trans. Am. Math. Soc.} {\bf {#1}} (19{#2}) #3}

\vglue 1.5truein
\begin{title}  
\righttext {MUTP/98/2}
\righttext{hep-th/9802029}
\centertext {\Bigfonts \bf Conformal anomaly in 2d}
\centertext{\Bigfonts \bf dilaton--scalar theory}
\vskip 20truept 
\centertext{J.S.Dowker\footnote{dowker@a3.ph.man.ac.uk}}
\vskip 7truept
\centertext{\it Department of Theoretical Physics,\\
The University of Manchester, Manchester, England}
\vskip10truept
\vskip 20truept
\centertext {Abstract}
\vskip10truept
\begin{narrow}
The discrepancy between the anomaly found by Bousso and Hawking and that of
other workers is explained by the omission of a zero mode contribution to
the effective action.
\end{narrow}
\vskip 5truept
\righttext {February 1998}
\vskip 60truept
\vfil
\end{title}
\pagenum=0
\section {\bf 1. Introduction}
A number of recent, and not so recent, papers have been concerned with the
conformal anomaly in the dilaton--scalar system in two-dimensional
gravity. This anomaly takes the general, local form 
$$
{T^\mu}_\mu=T(x)={1\over24\pi}(R-6\nabla^\mu\phi\nabla_\mu\phi
+\al\square\phi)
\eql{anom}$$
where $\phi$ is the dilaton field. The treatments agree on the first two
terms but the coefficient, $\al$, of the total divergence is subject to
some discussion. Firstly there is the question of whether the correct
two-dimensional reduction of the spherical four-dimensional theory has 
been taken. The reduction adopted by Elizalde, Naftulin and Odintsov, 
[\pref{ENO}], Mukhanov, Wipf and Zelnikov, 
[\pref{MWZ}], and by Kummer, Liebl and Vassilevich (KLV),
[\pref{KLV,KLV2}], produces $\al=6$. (This follows from the choice 
$\vphi=\psi=\phi$ in [\pref{KLV}]. See also Chiba and Siino, 
[\pref{CandS}]). Bousso and Hawking (BH), [\pref{BoandH}], effectively
choose a measure such that, in the notation of KLV, 
$\psi=0$ and $\vphi=\phi$. BH obtain the value $\al=-2$ while KLV's
formula gives $\al=4$. This latter value is also obtained by Ichinose,
[\pref{Ichinose}]. In this brief note we consider only this 
discrepancy since it seems to be a clear mathematical contradiction. 
The existence of a discrepancy, actually between the $\al=-2$ and $\al=6$ 
values, was early noted by Nojiri and Odintsov, [\pref{NandO}], who ascribed 
it entirely to total divergence ambiguities. For completeness, by repeating 
some standard material, we will here confirm the value $\al=4$ and then 
indicate where we think the calculation of BH breaks down.

\section{\bf 2. The dynamics and the anomaly}
In its simplest form the (matter) action adopted is, in 2d,
$$
S_m=-{1\over2}\int_\man \,e^{-2\phi}
\nabla^\mu f\,\nabla_\mu f\,\sqrt{g}d^2x
$$
where $f$ is the scalar matter field, with the corresponding field
operator
$$
A=e^{-2\phi}(-\square+2\nabla^\mu\phi\nabla_\mu).
\eql{op}$$

The most rapid method of finding the anomaly relies on its standard 
expression, $\ze(0,x)$, in terms of the local \zf\
associated with $A$, or, entirely equivalently, of the heat-kernel 
coefficient, $C^{(2)}_1(x)$, [\pref{DandC}]. To this end the operator 
$A$ is rewritten
$$
A=-e^{-2\phi}\big((\nabla^\mu-\nabla^\mu\phi)
(\nabla_\mu-\nabla_\mu\phi)+V\big)
$$
where 
$$V=\square\phi-\nabla^\mu\phi\nabla_\mu\phi.
$$

Introducing the auxiliary metric 
$$g'_{\mu\nu}=e^{2\phi}g_{\mu\nu}
$$
$A$ can also be written as
$$
A=-\big(({\nabla'}^\mu-{\nabla'}^\mu\phi)
({\nabla'}_\mu-{\nabla'}_\mu\phi)+V'\big)
$$
with
$$
V'=\square'\phi-{\nabla'}^\mu\phi{\nabla'}_\mu\phi=e^{-2\phi}\,V,
$$
where ${\nabla'}^\mu$ is ${\nabla'}_\mu,\,=\nabla_\mu$, raised by 
$g'^{\mu\nu}$.

When computing the eigenvalues of $A$, the scalar product of the $f$'s is
defined using the covariant measure of the $g$ metric. However, the trivial
Weyl potential, ${\nabla'}_\mu\phi$, can be removed by the gauge 
transformation, $f\to f'=\exp(-\phi)\,f$. The $f'$'s are 
normalized using the auxiliary metric, $g'$, and have the field operator $A'$
where
$$
A'=e^{-\phi}Ae^{\phi}=-\big(\square'+V'\big).
$$
The formal computation of $\ze(0)$ can thus proceed as for the standard
Laplacian by treating, temporarily, $g'$ as the metric. We will therefore 
find the integrated anomaly
$$
T=\ze(0)={1\over4\pi}\int\,C_1^{(2)}(g',x)\, \sqrt{g'}\,d^2x
$$
and can use the expression for $C_1$ derived many years ago,
$$
C^{(n)}_1(g',x)={R'\over6}+V',
\eql{cee1n}$$
where the coordinate system has been extended artificially to an 
$n$-dimensional one for later use.

The local trace anomaly, expressed as a density in the auxiliary metric is,
[\pref{DandC}], 
$$
T'(x)={1\over4\pi}C^{(2)}_1(g',x)={1\over4\pi}\bigg({R'\over6}+V'\bigg).
\eql{auxanom}$$
In order to obtain a density in the original metric, $g$, one simply
rewrites the $g'$ in (\peq{auxanom}) in terms of $g$ and removes the
resulting overall factor of $e^{-2\phi}$ to allow for the change in
the $\sqrt{g}$'s. As advertised we find,
$$
T(x)={1\over24\pi}(R-6\nabla^\mu\phi\nabla_\mu\phi+4\square\phi).
$$

Being based on standard techniques, this discussion adds nothing
material to the earlier treatments. However, as an amusing novelty, one 
can check the total derivative term, $\square'\phi$, in (\peq{cee1n}) in
the following way.

Instead of introducing the gauge potential we treat the operator $A$ as
it stands in (\peq{op}) and, further, work in $n$ dimensions. The idea here
is to use our previous technique, [\pref{Dowk}], of deriving the total 
derivative term in the {\it local} coefficient from the {\it integrated} 
coefficient (from which of course this term is absent).  

To save writing a lot of primes we replace $g'$ in $A$ by $g$ 
which should be thought of simply as a generic metric. 
(This is only a notational convenience for the purposes of this check.) 

Because $A$ is not conformally covariant in $n$
dimensions the calculation is not quite straightforward, but the necessary
formalism is available in [\pref{Dowk}]. The behaviour of $A$ under scale
changes $g_{\mu\nu}\to\overline g_{\mu\nu}=\la^2 g_{\mu\nu}$ is easily
determined to be
$$
(\overline A+\overline U)\overline f=\la^{-(n+2)/2}A f,\quad \overline f=
\la^{(2-n/2}f,
$$
where $\overline U$ measures the loss of conformal covariance
and equals, (\cf [\pref{Dowk}]),
$$
\overline U=(n-2)\overline\nabla^\mu\om\overline\nabla_\mu\phi+
\xi(n)(n-1)\big(2\la^{-1}\overline\square\la-(n-2)\overline\nabla^\mu
\om\,\overline\nabla_\mu\om\big)
$$
with $\xi(n)=(n-2)/4(n-1)$ and $\om=-\ln\la$. The second part of 
$\overline U$ is connected with the noncovariance of the Laplacian. 

Working around $\om=0$ (when $\overline U$ vanishes) and applying 
perturbation theory in $\overline U$ allows one to relate the 
relevant \zfs\ and thence the heat-kernel coefficients to obtain, 
[\pref{Dowk}],
$$\eqalign{
{1\over\sqrt{g}}{\de C_k^{(n)}[e^{-2\om}g]\over\de\om(x)}\bigg|_{\om=0}=&
-(n-2k)\,C_k^{(n)}(g,x)\cr
&+\bigg((n-2)\big(\square\phi+\nabla_\mu\phi\nabla^\mu\big)
+2(n-1)\xi(n)\,\square\bigg)C_{k-1}^{(n)}(g,x).\cr}
\eql{coeffvar}$$
We use this equation to find the local coefficient on the right from the
variation of the integrated one on the left.

As our application we set $k=1$. Using the fact that $C_0$ is the 
Weyl volume term, $C_0^{(n)}(g,x)=1$, we quickly find
$$
C^{(2)}_1(g,x)=\square\phi+
\lim_{n\to2}{1\over n-2}{1\over\sqrt{g}}{\de C_1^{(n)}
[e^{-2\om}g]\over\de\om(x)}\bigg|_{\om=0}.
\eql{cee1}$$

The point of this little exercise is simply to say that, assuming we know 
only the integrated $C_1^{(n)}$, 
$$
C_1^{(n)}[g]=\int\,\bigg({R\over6}-(\nabla\phi)^2\bigg)\,\sqrt{g}\,d^nx,
$$
then the variation and limit in (\peq{cee1}) easily yield
$$
C^{(2)}_1(g,x)=\square\phi+{R\over6}-(\nabla\phi)^2
$$
showing the resurrection of the total derivative contribution.

\section{\bf 3. Discussion}
We now turn to the question raised earlier concerning the origin of the 
discrepancy with the result of BH, [\pref{BoandH}]. 
The problem arises when BH assume, after their equn. (3.4), that the manifold
has the topology of the two-sphere, for then there is a zero mode of the
Laplacian and one cannot use the quoted Polyakov form for the 
effective action (equn. (3.3)). It is better to use the antisymmetrical 
cocycle function, $W[\overline g,g]\sim W[\overline g]-W[g]$, for the 
conformal change $g\to\overline g$.  

As we have shown, [\pref{Dowk2}], because of the zero 
mode, apart from the standard contribution, there is an additional term of 
the form
$$
\De W[\overline g,g]=
{1\over48\pi |\man|}\int\ln\bigg({g\over\overline g}\bigg)\sqrt{g}\, d^2x\,
\int R\sqrt{g}\, d^2x
$$
where $|\man|=|\man(g)|$ is the two-surface area.

Computing this for the uniform rescaling, $\overline g_{\mu\nu}=
\exp(2\phi_c)g_{\mu\nu}$, yields an extra contribution which cancels the
change in the effective action used by BH -- the last term in equn. (3.5).
(This must be so for consistency and is the whole point of [\pref{Dowk2}].) 
If we carried on with the analysis as in BH, then we would conclude that 
$q_3,\,=\al/24\pi$, were zero. 

One way of partially retrieving the situation is to use the cocycle 
function, $W[\overline g,g]$, (as we 
should). Then the last term in equn. (3.3) is replaced by
$$
{1\over2}q_3\int \big(\overline R\sqrt{\overline g}-R\sqrt{g}\,
\big)\phi\,d^2x
$$
which vanishes when $\phi$ is uniform by topological invariance, but which 
still has the required variation and everything is consistent. However it 
is not then possible to deduce 
the value of $q_3$ in a simple way, at least not by conformal transformations
in two dimensions alone.

If we wish to avoid a zero mode, then it is necessary to include 
boundary terms in the effective action, 
{\it when this last is being evaluated}. In any case, zero mode or boundary,
the additional contributions remove any discrepancies and also, 
incidentally, render nugatory the specific criticisms by Nojiri and 
Odintsov, [\pref{NandO}], of Bousso and Hawking's choice of term in the 
effective action. The problem is not so much the ambiguity in this term, 
rather it is its incorrect behaviour for uniform $\phi$.

\section{\bf References}
\vskip 5truept
\begin{putreferences}
\ref{CandT}{Copeland,E. and Toms,D.J.\np{255}{85}{201};\cqg{4}{86}{1357}}
\ref{BKD}{Bordag,M., K.Kirsten,K. and Dowker,J.S.: \cmp{}{96}{}in the press.}
\ref{BBG}{Blazik,N., Bokan,N. and Gilkey,P.B.: Spectral geometry of the 
form valued Laplacian for manifolds with boundary \ijpam{23}{92}{103-120}}
\ref{CandK}{Cognola,G. and Kirsten,K. \cqg{13}{96}{633}.}
\ref{ELV2}{Elizalde, E., Lygren, M. and Vassilevich, D.V. : Zeta function 
for the laplace operator acting on forms in a ball with gauge boundary 
conditions. hep-th/}
\ref{ELV1}{Elizalde, E., Lygren, M. and Vassilevich, D.V. : Antisymmetric
tensor fields on spheres: functional determinants and non-local
counterterms, \jmp{}{96}{} to appear. hep-th/ 9602113}
\ref{ENO}{E.Elizalde, S.Naftulin and S.D.Odintsov \prD{49}{94}{2853}.}
\ref{CandH2}{Camporesi,R. and Higuchi, A.: Plancherel measure for $p$-forms
in real hyperbolic space preprint 1992} 
\ref{APS}{Atiyah,M.F., V.K.Patodi and I.M.Singer: Spectral asymmetry and 
Riemannian geometry \mpcps{77}{75}{43}.}
\ref{AandT}{Awada,M.A. and D.J.Toms: Induced gravitational and gauge-field 
actions from quantised matter fields in non-abelian Kaluza-Klein thory 
\np{245}{84}{161}.}
\ref{BandI}{Baacke,J. and Y.Igarishi: Casimir energy of confined massive 
quarks \prD{27}{83}{460}.}
\ref{Barnesa}{E.W.Barnes {\it Trans. Camb. Phil. Soc.} {\bf 19} (1903) 
374.}
\ref{Barnesb}{E.W.Barnes {\it Trans. Camb. Phil. Soc.} {\bf 19} (1903) 
426.}
\ref{Barv}{Barvinsky,A.O. Yu.A.Kamenshchik and I.P.Karmazin: One-loop 
quantum cosmology \aop {219}{92}{201}.}
\ref{BandM}{Beers,B.L. and Millman, R.S. :The spectra of the Laplace-Beltrami
operator on compact, semisimple Lie groups. \ajm{99}{75}{801-807}.}
\ref{BandH}{Bender,C.M. and P.Hays: Zero point energy of fields in a 
confined 
volume \prD{14}{76}{2622}.}
\ref{BEK}{Bordag,M., E.Elizalde and K.Kirsten {\it Heat kernel 
coefficients of the Laplace operator on the D-dimensional ball}, 
\jmp{37}{96}{895}.}
\ref{BGKE}{Bordag,M., B.Geyer, K.Kirsten and E.Elizalde, {\it Zeta function
determinant of the Laplace operator on the D-dimensional ball}, 
\cmp{179}{96}{215}.}
\ref{Branson}{Branson,T.P.: Conformally covariant equations on differential
forms \cpde{7}{82}{393-431}.}
\ref{BandG2}{Branson,T.P. and P.B.Gilkey {\it Comm. Partial Diff. Eqns.}
{\bf 15} (1990) 245.}
\ref{BoandH}{R.Bousso and S.W.Hawking \prD{56}{97}{7788}.}
\ref{BGV}{Branson,T.P., P.B.Gilkey and D.V.Vassilevich {\it The Asymptotics
of the Laplacian on a manifold with boundary} II, hep-th/9504029.}
\ref{CandH}{Camporesi,R. and A.Higuchi {\it On the eigenfunctions of the 
Dirac operator on spheres and real hyperbolic spaces}, gr-qc/9505009.}
\ref{ChandD}{Peter Chang and J.S.Dowker \np{395}{93}{407}.}
\ref{cheeg1}{Cheeger, J.: Spectral Geometry of Singular Riemannian Spaces.
\jdg {18}{83}{575}.}
\ref{cheeg2}{Cheeger,J.: Hodge theory of complex cones {\it Ast\'erisque} 
101-102.}
\ref{Chou}{Chou,A.W.: The Dirac operator on spaces with conical singularities
and positive scalar curvature, \tams{289}{85}{1-40}.}
\ref{CandS}{T.Chiba and M.Siino \mplA{12}{97}{709}.}
\ref{DandH}{D'Eath,P.D. and J.J.Halliwell: Fermions in quantum cosmology 
\prD{35}{87}{1100}.}
\ref{cheeg3}{Cheeger,J.:Analytic torsion and the heat equation. \aom{109}
{79}{259-322}.}
\ref{DandE}{D'Eath,P.D. and G.V.M.Esposito: Local boundary conditions for 
Dirac operator and one-loop quantum cosmology \prD{43}{91}{3234}.}
\ref{DandE2}{D'Eath,P.D. and G.V.M.Esposito: Spectral boundary conditions in 
one-loop quantum cosmology \prD{44}{91}{1713}.}
\ref{Dow1}{J.S.Dowker \cmp{162}{94}{633}.}
\ref{Dow8}{Dowker,J.S. {\it Robin conditions on the Euclidean ball} 
MUTP/95/7; hep-th\break/9506042. {\it Class. Quant.Grav.} to be published.}
\ref{Dow9}{Dowker,J.S. {\it Oddball determinants} MUTP/95/12; 
hep-th/9507096.}
\ref{Dow10}{Dowker,J.S. {\it Spin on the 4-ball}, 
hep-th/9508082, {\it Phys. Lett. B}, to be published.}
\ref{Dowk}{J.S.Dowker \prD{39}{89}{1235}.}
\ref{Dowk2}{J.S.Dowker \cqg{11}{94}{L7}.}
\ref{DandC}{J.S.Dowker and R.Critchley \prD{16}{77}{3390}.}
\ref{DandA2}{Dowker,J.S. and J.S.Apps, {\it Functional determinants on 
certain domains}. To appear in the Proceedings of the 6th Moscow Quantum 
Gravity Seminar held in Moscow, June 1995; hep-th/9506204.}
\ref{DABK}{Dowker,J.S., Apps,J.S., Bordag,M. and Kirsten,K.: Spectral 
invariants for the Dirac equation with various boundary conditions \cqg{}{}{}
to be published, hep-th/9511060.}
\ref{Kam2}{Esposito,G., A.Y.Kamenshchik, I.V.Mishakov and G.Pollifrone: 
Gravitons in one-loop quantum cosmology \prD{50}{94}{6329}; 
\prD{52}{95}{3457}.}
\ref{Esposito}{Esposito,G. {\it Quantum Gravity, Quantum Cosmology and 
Lorentzian Geometries}, Lecture Notes in Physics, Monographs, Vol. m12, 
Springer-Verlag, Berlin 1994.}
\ref{Esposito2}{Esposito,G. {\it Nonlocal properties in Euclidean Quantum
Gravity}. To appear in Proceedings of 3rd. Workshop on Quantum Field Theory
under the Influence of External Conditions, Leipzig, September 1995; 
gr-qc/9508056.}
\ref{ETP}{Esposito,G., H.A.Morales-T\'ecotl and L.O.Pimentel {\it Essential
self-adjointness in one-loop quantum cosmology}, gr-qc/9510020.}
\ref{FORW}{Forgacs,P., L.O'Raifeartaigh and A.Wipf: Scattering theory, U(1) 
anomaly and index theorems for compact and non-compact manifolds 
\np{293}{87}{559}.}
\ref{GandM}{Gallot S. and Meyer,D. : Op\'erateur de coubure et Laplacian
des formes diff\'eren-\break tielles d'une vari\'et\'e riemannienne 
\jmpa{54}{1975}
{289}.}
\ref{Gilkey1}{Gilkey, P.B, Invariance theory, the heat equation and the
Atiyah-Singer index theorem, 2nd. Edn., CTC Press, Boca Raton 1995.}
\ref{Gilkey2}{Gilkey,P.B.:On the index of geometric operators for Riemannian 
manifolds with boundary \aim{102}{93}{129}.}
\ref{Gilkey3}{Gilkey,P.B.: The boundary integrand in the formula for the 
signature and Euler characteristic of a manifold with boundary 
\aim{15}{75}{334}.}
\ref{Grubb}{Grubb,G. {\it Comm. Partial Diff. Eqns.} {\bf 17} (1992) 
2031.}
\ref{GandS1}{Grubb,G. and R.T.Seeley \cras{317}{1993}{1124}; \invm{121}{95}
{481}.}
\ref{Ichinose}{S.Ichinose {\it Weyl anomaly of 2D Dilaton-Scalar and 
Hermiticity of System Operator}, hep-th/9707025.}
\ref{IandT}{Ikeda,A. and Taniguchi,Y.:Spectra and eigenforms of the Laplacian
on $S^n$ and $P^n(C)$. \ojm{15}{78}{515-546}.}
\ref{IandK}{Iwasaki,I. and Katase,K. :On the spectra of Laplace operator
on $\La^*(S^n)$ \pja{55}{79}{141}.}
\ref{JandK}{Jaroszewicz,T. and P.S.Kurzepa: Polyakov spin factors and 
Laplacians on homogeneous spaces \aop{213}{92}{135}.}
\ref{Kam}{Kamenshchik,Yu.A. and I.V.Mishakov: Fermions in one-loop quantum 
cosmology \prD{47}{93}{1380}.}
\ref{KandM}{Kamenshchik,Yu.A. and I.V.Mishakov: Zeta function technique for
quantum cosmology {\it Int. J. Mod. Phys.} {\bf A7} (1992) 3265.}
\ref{KandC}{Kirsten,K. and G.Cognola, {\it Heat-kernel coefficients and 
functional determinants for higher spin fields on the ball} UTF354. Aug. 
1995, hep-th/9508088.}
\ref{KLV}{W.Kummer, H.Liebl and D.V.Vassilevich \mplA {12}{97}{2683}.}
\ref{KLV2}{W.Kummer, H.Liebl and D.V.Vassilevich {\it Comment on ``Trace
anomaly of dilaton coupled scalars in two dimensions''}, hep-th/9801122.}
\ref{Levitin}{Levitin,M. {\it Dirichlet and Neumann invariants for Euclidean
balls}, {\it Diff. Geom. and its Appl.}, to be published.}
\ref{Luck}{Luckock,H.C.: Mixed boundary conditions in quantum field theory 
\jmp{32}{91}{1755}.}
\ref{MandL}{Luckock,H.C. and I.G.Moss: The quantum geometry of random 
surfaces and spinning strings \cqg{6}{89}{1993}.}
\ref{Ma}{Ma,Z.Q.: Axial anomaly and index theorem for a two-dimensional disc 
with boundary \jpa{19}{86}{L317}.}
\ref{Mcav}{McAvity,D.M.: Heat-kernel asymptotics for mixed boundary 
conditions \cqg{9}{92}{1983}.}
\ref{MandV}{Marachevsky,V.N. and D.V.Vassilevich {\it Diffeomorphism
invariant eigenvalue \break problem for metric perturbations in a bounded 
region}, SPbU-IP-95, \break gr-qc/9509051.}
\ref{Milton}{Milton,K.A.: Zero point energy of confined fermions 
\prD{22}{80}{1444}.}
\ref{MandS}{Mishchenko,A.V. and Yu.A.Sitenko: Spectral boundary conditions 
and index theorem for two-dimensional manifolds with boundary 
\aop{218}{92}{199}.}
\ref{Moss}{Moss,I.G.: Boundary terms in the heat-kernel expansion 
\cqg{6}{89}{759}.}
\ref{MandP}{Moss,I.G. and S.J.Poletti: Conformal anomaly on an Einstein 
space with boundary \pl{B333}{94}{326}.}
\ref{MandP2}{Moss,I.G. and S.J.Poletti \np{341}{90}{155}.}
\ref{MWZ}{V.Mukhanov, A. Wipf and A. Zelnikov \pl{B332}{94}{283}.}
\ref{NandO}{S.Nojiri and S.D.Odintsov \mplA{12}{97}{2083}.}
\ref{NandS}{Niemi,A.J. and G.W.Semenoff: Index theorem on open infinite 
manifolds \np {269}{86}{131}.}
\ref{NandT}{Ninomiya,M. and C.I.Tan: Axial anomaly and index thorem for 
manifolds with boundary \np{245}{85}{199}.}
\ref{norlund2}{N\"orlund~N. E.:M\'emoire sur les polynomes de Bernoulli.
\am {4}{121} {1922}.}
\ref{Poletti}{Poletti,S.J. \pl{B249}{90}{355}.}
\ref{RandS}{R\"omer,H. and P.B.Schroer \pl{21}{77}{182}.}
\ref{Trautman}{Trautman,A.: Spinors and Dirac operators on hypersurfaces 
\jmp{33}{92}{4011}.}
\ref{Vass}{Vassilevich,D.V.{\it Vector fields on a disk with mixed 
boundary conditions} gr-qc /9404052.}

\end{putreferences}

\bye